%% file: ms.tex
\documentclass[structabstract]{aa}  
\usepackage{graphicx}
\usepackage{txfonts}
\usepackage{natbib}
\usepackage{amssymb}
\begin{document}
   \title{Looking for outflow and infall signatures in high mass star forming regions }
    \titlerunning{Infall and outflow in high mass SFRs}

   \author{P. D. Klaassen\inst{1,2}
          \and
          L. Testi\inst{1,3} 
          \and
          H. Beuther\inst{4}         }

   \institute{European Southern Observatory, Karl Schwarzschild Str 2, 85748 Garching, Germany \\           
      \email{klaassen@strw.leidenuniv.nl}
         \and
         Leiden Observatory, Leiden University, PO Box 9513, 2300 RA, Leiden, The Netherlands
         \and
         INAF-Osservatorio Astrofisico di Arcetri, Largo E. Fermi 5, 50125 Firenze, Italy
         \and
             Max Planck Institute for Astronomy, K\"onigstuhl 17, 69117 Heidelberg, Germany             }

   \date{Received }

 
  \abstract
   {Many physical parameters change with time in star forming regions. Here we attempt to correlate changes in infall and outflow motions in high mass star forming regions with evolutionary stage using JCMT observations. }
   {From a sample of 45 high mass star forming regions in three phases of evolution, we investigate the presence of established infall and outflow tracers to determine whether there are any trends attributable to the age of the source.}
   {We obtained JCMT observations of HCO$^+$/H$^{13}$CO$^+$ J=4-3 to trace large scale infall, and SiO J=8-7 to trace recent outflow activity.  We compare the infall and outflow detections to the evolutionary stage of the host source (high mass protostellar objects, hypercompact HII regions and ultracompact HII regions). We also note that the integrated intensity of SiO varies with the full width at half maximum of the H$^{13}$CO$^+$.}
   {We find a surprising lack of SiO detections in the middle stage (Hypercompact HII regions), which may be due to an observational bias.  When SiO is detected, we find that the integrated intensity of the line increases with evolutionary stage.  We also note that all of the sources with infall signatures onto Ultracompact HII regions have corresponding outflow signatures as well.}
 {}


   \maketitle
%

\section{Introduction}

High mass stars are a vital player in the evolution of galaxies. During their formation they inject energy into their surroundings in the form of outflows and ionizing photons. As they evolve their stellar winds continue to stir turbulence in the surrounding medium, and when they explode as supernovae they additionally enhance the metallicity of their environs.

The processes involved in this earliest stage, the formation of high mass stars is not nearly as well constrained as are the mechanisms responsible for lower mass stars. The pre main sequence evolution of low mass stars \citep[i.e.][]{shu87,andre93} has been well constrained and statistically based lifetimes of the stages have come about, in part, due to observing large numbers of individual sources \citep[i.e][]{Enoch09,Spezzi08}.

Similar studies for the early evolution of high mass stars (M$\gtrsim 8$ M$_\odot$) have proven problematic due to limiting factors such as the large average distances to high mass star forming regions and the clustered nature of their formation.  However studies such as \citet{Beltran11} suggest high mass stars are likely to form though disk mediated accretion in a similar but scaled up version of lower mass star formation. An evolutionary sequence is developing, however progress is hampered by a lack of large sample surveys of the molecular gas dynamics in these region \citep[see, for example][]{ZY07,Beuther07}. Evolution in low mass systems is often characterized by changes in the spectral energy distributions (SEDs). This is not as easily done in high mass systems \citep{Molinari08}, which is why studying the gas dynamics is very important.

In broad terms, the evolution of massive protostars begins in an infrared dark cloud \citep[IRDC, i.e.][]{Egan98}. The central condensation then begins heating its environment, at which point the core starts emitting at IR wavelengths and it becomes a high mass protostellar object \citep[HMPO, i.e.][]{Sridharan02} within a hot core.  The protostar continues to gain mass and heat its environment eventually forming a hypercompact (HC) HII region \citep[i.e.][]{Keto07} which then grows into an ultracompact (UC) HII region \citep[i.e.][]{WC89}.  This sequence has been shown graphically in \citet{Zapata10}.  The goal of this paper is to follow the larger scale (that observable with a single dish) outflow and infall structures surrounding sources in the later three of these four evolutionary stages (HMPO, HCHII and UCHII region), and determine whether any evolutionary trends are distinguishable.

In the case of low mass star formation, \citet{andre93} conclude that outflow motions can be seen quite early in the Class 0 phase, and \citet{Bontemps96} state that the outflows from Class 0 sources are more powerful than those from Class I sources at the same luminosity.  They also suggest that all deeply embedded low mass protostars should have outflows. As for infall in the low mass regime, inverse P-Cygni profiles indicative of infall have been seen for a decade now \citep{difrancesco01,Zapata08,Furuya11}, and some observations even suggest that infall is seen perpendicular to the outflow direction \citep{Arce04}. That these trends can be applied to higher mass protostars is starting to be explored \citep[i.e.][]{LS11}.

Here we present HCO$^+$ and H$^{13}$CO$^+$ (J=4-3) observations to study the bulk infall signatures in high mass star forming regions.  We also present SiO (J=8-7) observations to trace recent outflow activity from the same sources. We have collected a sample of 45 high mass star forming regions in three evolutionary stages: HMPOs, and sources with HC and UCHII regions.  In Section \ref{sec:observations} we present our observations and in Section \ref{sec:method} we present our methodology.  In Section \ref{sec:results} we present our results, which we discuss in Section \ref{sec:discussion} where we also conclude.


\section{Observations}
\label{sec:observations}

\citet{KW07,KW08} observed 24 high mass star forming regions with HII region in the hypercompact (2) and ultracompact (22) stages of evolution with Receiver B and HARP-B on the James Clerk Maxwell Telescope (JCMT)\footnote{The James Clerk Maxwell Telescope is operated by The Joint Astronomy Centre on behalf of the Science and Technology Facilities Council of the United Kingdom, the Netherlands Organisation for Scientific Research, and the National Research Council of Canada.}. They observed HCO$^+$/H$^{13}$CO$^+$ (J=4-3) to detect infall signature, and SiO (J=8-7) to trace current outflow motions.

Here we are adding observations of earlier stage sources: 12 HMPOs, and 9 hypercompact HII regions in the same tracers mapped with HARP-B in August 2010.  These observations (Project M10BI03) were taken in weather bands 2 and 3 with a goal rms noise level of 0.1 K in the T$_{\rm A}^*$ temperature scale for the 356 GHz observations (HCO$^+$) and 0.07 K for the 347 GHz observations (H$^{13}$CO$^+$/SiO). The velocity resolutions were 0.42 km s$^{-1}$ and 0.82 km s$^{-1}$ respectively for the 356 and 347 GHz observations. The beam size is 15$''$ at these frequencies.  Position switched 16$\times$16 pixel jiggle maps were created for each source and the rms limits achieved for each map are given in Table \ref{tab:observations}.  The central spectra for each source are shown in Figure \ref{fig:spectra}. Note that in Figure \ref{fig:spectra}, the H$^{13}$CO$^+$ intensities have been scaled up by a factor of four, and the SiO intensities by a factor of 16.

The new observations were reduced using the JCMT pipeline. For the 347 GHz observations of G213.88-11.84, the system was set to `tracking' to remove an angle offset between multiple scans.  Once processed, the datacubes were exported to MIRIAD format for further analysis.

\subsection{Source Selection}

The HC and UCHII regions studied in \citep{KW07,KW08} were selected based on having been part of the \citep{WC89} or \citep{Kurtz94} surveys of HII regions associated with high mass star formation, and in either \citet{P92} or \citet{Hunter97} to show they had outflow motions.

In order to probe earlier stages of high mass star formation, we selected nine more sources from \citet{Kurtz94} in which the HII regions was smaller than 0.01 pc, more characteristic of HCHII regions than ultracompact ones and were previously known to have outflows.  The HII region sizes were determined by taking the observed source diameters from Table 3 of \citet{Kurtz94} and the distances they give in their Table 2. The HMPOs in this study come from the \citet{Sridharan02} sample of sources. More specifically, we chose sources which were observed by \citet{Beuther02} in CO. For simplicity we excluded sources which had multiple velocity components  to remove regions which may have multiple outflows, as well as those with the narrowest line widths, to maximize the likelihood of detecting SiO. In summary, the determining factors for the classification in the threes stages were: inclusion in \citet{Beuther02} (HMPOs); HII region smaller than 0.01 pc (HCHII regions); or HII regions larger than 0.01 pc (UCHII regions).

\input{tab1.tab}

\input{tab2.tab}

\input{tab3.tab}


\section{Methodology}
\label{sec:method}

\subsection{Positions for Profiles}

It was noted that for a number of sources (11) in this study, the position of the continuum peak and the position where the optically thin H$^{13}$CO$^+$ is brightest can be offset from each other even at the 15$''$ resolution of the JCMT at 350 GHz. When present, molecular peaks were generally only 15$''$ (one beam) from the continuum peak, in random directions. This was also noted for a number of the HMPOs presented in \citet{Beuther02} at the 11$''$ resolution of their IRAM observations.

For those sources with emission maps (all of the new data, and those presented in \citet{KW08}) we are able to compare infall and outflow statistics both at the continuum peak position or the molecular peak position.

For the HMPOs, the position of the continuum peak is taken from Table 2 of \citet{Beuther02a} and for the sources from \citet{WC89} and \citet{Kurtz94}, the continuum peak positions were taken from their Tables 4 and 2 (respectively).  Below, if the positions are not explicitly stated, properties are derived from the continuum peak and not the (possibly) offset molecular peak. This is to keep the mapped data consistent with the single pixel data, which was consistently taken at the continuum peak.  Whether the molecular line and continuum peaks are offset from each other is noted in the last column of Table \ref{tab:detections}

\subsection{Detecting Infall Signatures with HCO$^+$}

HCO$^+$ is an excellent tracer of high density gas, like that found in high mass star forming regions. It has been used by many authors to trace large scale infall in both low mass \citep{mardones97,gregersen97} and high mass \citep{Fuller05,KW07,KW08} star forming regions.  CS is another good tracer of high density gas however \citet{Sun09} suggest that HCO$^+$ is a better tracer of infall motions \citep[see also][]{Tsamis08}.  

The less optically thick H$^{13}$CO$^+$ isotopologue was also observed in order to distinguish between infall signatures and multiple line of sight components, as described below.

Double peaked or asymmetric HCO$^+$ line profiles in which the brightest emission is blueward of the source rest velocity can be interpreted as due to infall if the H$^{13}$CO$^+$ is single peaked at the rest velocity.  If the optically thin H$^{13}$CO$^+$ also has a double peaked profile, the double peak in both lines is likely due to multiple components along the line of sight. \citet{Churchwell10} describe the different mechanisms that could produce a double peak in an HCO$^+$ line profile towards massive star forming regions, and conclude that self absorption best fits observed line profiles. The blue or red skew of the absorption profile can then be used to distinguish infall (blue) or outflow (red).

In our study we find 23 sources (51\%) with blue asymmetries, suggestive of infall in HCO$^+$; 3/12 in the HMPO phase, 8/11 in the HCHII region phase, and 12/22 in the UCHII region phase, see Table \ref{tab:stats} and Figure \ref{fig:frac_stage}.

We define an infall detection for sources with double peaked HCO$^+$ profiles (and single peaked H$^{13}$CO$^+$) such as IRAS18182-1433, and  for sources with asymmetrically blue profiles containing red shoulders like G81.68. Of the 14 sources in the mapped data with infall signatures 6(7) of these sources have double peaked profiles at the continuum (molecular) peak positions.  For all sources with double peaked HCO$^+$ profiles with stronger blue peaks than red, we were able to calculate infall velocities and mass infall rates. We determined the velocities and intensities of the blue and red peaks, and that of the absorption feature at its minimum.  We also determined the FWHM of the optically thin H$^{13}$CO$^+$ and input all of these quantities into Equation 9 of \citet{Myers96} to determine the infall velocity.  To determine the mass infall rate, we used Equation 3 of \citet{KW07} where $\dot{M}=(4/3)\pi n_{\rm H_2}\mu m_{\rm H}r^2V_{\rm in}$ is the mass infall rate, $\mu$ is the mean molecular weight (2.35), $m_{\rm H}$ is the mass of Hydrogen, $r$ is the radius of the emitting region (assumed here to be the beam radius), $V_{\rm in}$ is the infall velocity, and $n_{\rm H_2}$ is the ambient density taken from previous studies \citep[i.e.][]{Beuther02a,P97,Shirley03}. For sources where no ambient density was found, the average of the ambient densities from the other sources ($n_{\rm H_2}=10^6$) was used.  These values are shown in Table \ref{tab:infall_vel}. For the source with a `shoulder' infall signature, we were unable to calculate these properties. Note that as the mass infall rate scales with the square of the infalling radius, the effects of beam dilution (source smaller than the beam) are likely to be over estimating the mass infall rate.

Our HMPOs are drawn from the same sample as those of \citet{Fuller05}. Of our 12 HMPOs, eight overlap with those in their sample which have significant line asymmetries (those listed in their Table 9).  In each case, our HCO$^+$ line asymmetries are consistent with theirs, save for one source IRAS22134+5834, where we detect shoulder emission. In the J=1-0 and J=3-2 transitions of HCO$^+$ for this source, they report red asymmetries and in J=4-3 they do not find any significant line asymmetry. They also note that the H$_2$CO towards this source shows a blue asymmetry.  In their total sample of 78 sources, they suggest 22 are infall candidates.  Their detection rate of 28\% is consistent with our detection rate of 25\% in 12 sources.

Twelve of our sources overlap with the HCO+ (J=3-2) study of \citet{Reiter11b}. For 11 of these sources, our line profiles (blue vs. red asymmetry) are consistent. For the other source (CepA), we find that our line profiles are consistent with the HCN profiles shown in \citet{Wu10} and \citet{Wu03} and the HCO$^+$ (J=1-0) profile in \citet{Sun09}.  \citet{Reiter11b} suggest that there is outflow contamination in their 30$''$ beam for CepA. 

The study of \citet{Churchwell10} observed a number of HII regions in HCO$^+$ (J=3-2,2-1). Nine of their sources overlap with our study.  We find that for all nine sources, whether or not we detect an infall signature is consistent with their findings.  This includes G33.13, where we both find red peaked profiles (which indicates outflow). They note that of their 24 HII regions (both ultracompact and hyper compact), they only detect infall signatures in eight sources (33\%).  For our 33 HII regions, we detect infall in 20 (61\%).

There are also nine sources in our survey that overlap with the study of \citet{LS10}. We compare our infall detections with their detections in HCO$^+$ J=1-0, and find we agree for six out of nine sources.  The sources for which we disagree are G192.58, G192.6 and IRAS23151. For each of these three sources, we detect infall in regions where they do not.  We suggest that our detections are partly due to our higher resolution (15$''$ vs. 29$''$), and because at earlier times, the J=4-3 transition is likely a better asymmetry tracer than J=1-0 \citep[see Figure 8 of][]{Tsamis08}.  This statement is supported by comparison with the HCO$^+$ J=1-0 observations of \citet{Purcell06}. Many more of their HCO$^+$ lines were double peaked, and that these asymmetries are equally split between red and blue.  Our asymmetries are skewed towards blue asymmetries.  The higher energy transition does not self absorb as readily, and stronger velocity gradients maybe required to show an infall signature.

\input{tab5.tab}

\subsection{Detecting Outflow Signatures with SiO}

As stated in Section \ref{sec:observations}, all of the sources in this survey were selected based on previous outflow signatures having been detected (in, for instance CO or CS).  Here, to classify sources as having active outflows, we require a detection of SiO J=8-7.   SiO is formed in shocked regions as Si is liberated from dust grains and joins with O \citep{Schilke97,Caselli97}. The Si then evolves out of species observable from the ground in $\sim 10^4$ yr {\citep{Pineau97}, and therefore SiO is an excellent tracer of recent shock activity. Because CO is a bulk gas tracer, and not a specific shock tracer like SiO, it gives a broader sense of the outflow history of the region. Because SiO is short lived, it traces active outflows.

To systematically determine whether SiO is detected we determined the width of the HCO$^+$ line at the 3 $\sigma$ level. This width and these velocity limits were used to create a zeroth moment map of SiO. If the intensity at the peak pixel (whether continuum or molecular peak) was greater than 3 $\sigma$ in that velocity interval, we claimed an SiO detection.

SiO was detected at the continuum peak in 23 sources; 6 in the HMPO phase, 2 in the HCHII region phase, and 15 in the UCHII region phase (see Table \ref{tab:stats} and Figure \ref{fig:frac_stage}). For sources with maps, where we could distinguish between the continuum and molecular peak, we found 4 more detections (1 HMPO, 2 HCHII and 1 UCHII region), bringing the total number of sources with SiO detections to 27 (60\%).  In no source where SiO was detected at the continuum peak was the signal lost at the molecular peak.

We note that the SiO detection rate for the J=2-1 and J=3-2 observations of \citet{LS11} are much higher (88\%).  The sources in \citet{LS11}  were taken from \citet{Rathborne06}, \citet{Beuther02}, \citet{Faundez04}, \citet{Hill05}, \citet{Hofner96}, \citet{Walsh98}, and \citet{WC89}. Reclassifying their IR lound and dark sources into our nomenclature, the \citet{Rathborne06} sources became Infrared Dark Clouds (IRDCs), The \citet{Beuther02} and \citet{Faundez04} sources became HMPOs, and the remaining sources were classified as either HC or UCHII regions depending on the size of the HII region taken from the literature. Without taking into consideration the excitation of the different transitions or the different beam sizes of the observations, but purely looking at whether these 2-1 and 3-2 detections would have been made at the sensitivity limits of our observations, we find that only  of order 55-60\% of their sources would be detected in our survey.    We are thus likely missing a few detections due to our sensitivity limits.  We note that increasing their rms levels to match ours, the fractional detection of SiO in HCHII regions drops much lower than it does for the HMPOs and the UCHII regions. This suggests that perhaps our lack of detection of SiO in HCHII regions maybe an observational bias (see Figure \ref{fig:comparison}).

 With regards to the excitation of SiO J=8-7 in our sources, we find very few sources with extended SiO emission.  We compared our SiO integrated intensities to those of the J=2-1 and J=3-2 transitions observed in \citet{LS11}.  We find a line ratio (8-7)/(3-2) $<$ 1 for all sources in both surveys which are detected in SiO.Comparing this result to the models of \citet{Gusdorf08}, assuming an ambient density of 10$^6$ cm$^{-3}$, we suggest low shock velocities ( $<$ 30 km s$^{-1}$) in these sources.  As in \citet{Leurini11c}, we find that SiO shock models developed for low mass star forming regions are consistent with our high mass star forming region results as well.


\section{Results}
\label{sec:results}

In Figure \ref{fig:averages} we present a number of source properties averaged by the presumed evolutionary stage of the source.  These properties include the FWHM of the H$^{13}$CO$^+$, the distance to the sources, the far infrared Luminosity of the source, and the integrated intensity of the SiO lines.  Note that prior to averaging, G5.89 was removed from the sample, as the properties of this source are much greater than 3$\sigma$ away from the averages and can skew our results.

The FWHM of the H$^{13}$CO$^+$, a measure of the turbulence in our observed regions does not seem to correlated with source evolutionary stage([stage] =0.5$\pm$0.9[H$^{13}$CO$^+$ FWHM]+3$\pm$2). The averaged trend in Figure \ref{fig:averages}  suggests a possible weak correlation, however the correlation coefficient is 0.025 when the data is not averaged (as stated in the figure caption). This suggests that the FWHM can be used  for determining source properties independent of its age.

We do appear to have a slight distance bias in our survey sources, with older sources being, on average, slightly further away. The correlation coefficient of the averaged data is 0.99, while the coefficient of the un-averaged data is only 0.25.  The distance differences are not likely to be significant in our analysis ([stage] = 1$\pm$[D]+2$\pm$4).

The FIR luminosities of our sources, taken from \citet{WC89,Kurtz94,Beuther02,Mateen06,Walsh97} are the IR fluxes from IRAS converted into total luminosities for the sources at the distances listed in those papers. The FIR luminosity can be taken as a tracer of the total luminosity of the source if the distance is known \citep{WC89}. Thus, the FIR luminosity can be used to determine the spectral type of the embedded protostar. On average the FIR luminosity does not seem to correlate with evolutionary stage ([stage]=0.71$\pm$0.41[Log L$_{\rm FIR}$]+3.35$\pm$0.83), with a correlation coefficient of 0.74 for three points.  When the individual FIR luminosities are considered (and not the average for the stage), the coefficient drops to 0.36. The slope and intercept of the least squares fits to the data remain the same between the averaged and un-averaged data. We thus suggest that the FIR luminosity does not correlate with evolutionary stage.  Neither the SIO integrated intensity nor the H$^{13}$CO$^+$ FWHM correlate with the FIR luminosity. The scaling relation for SIO integrated intensity (y) vs. FIR luminosity (x) is y=1.6$\pm$0.8x-4$\pm$4, while for H$^{13}$CO$^+$ (y) vs. FIR luminosity (x) is y=0.9$\pm$0.2-0.7$\pm$1.2 (with correlation coefficients of 0.36 and 0.5, respectively), nor does the SiO integrated intensity correlate with distance to source (correlation coefficient of 0.3).  However, the H$^{13}$CO$^+$ line width does correlate with distance, as described further in Section \ref{subsec:outflow}.

\input{tab4.tab}

\subsection{Infall}
\label{sec:infall}

We find that our detections of infall signatures drastically changes from the HMPO stage (where we only detect infall in 25\% of our sources) to the HCHII region phase, were we detect infall in 72\% of sources.  This result, is surprising given that the younger sources should be actively accreting material onto the central source.  However, if we consider that the infall may start at the center of the clump and the region undergoing infall expands outwards, then the infall area for a younger source is smaller, and thus more likely to be beam diluted.  Studies of the line asymmetry arising from infall \citep[i.e][]{Devries05,Tsamis08} suggest that infall signatures are easiest to distinguish when the infall region is best matched to the observing beam. 
For the HMPOs in our sample, the infall signatures are not preferentially seen in the closer sources. The average distance to a source with an infall detection is 4.6 kpc, while the average distance to sources without is 4.1 kpc. Thus, we suggest that the lack of detections is not a distance bias, but based on a general beam dilution of the signal.

Assuming the infall is occurring from the inside outward, then in the earlier stages, the infall region is much smaller than our observing beam, and thus the signal is beam diluted.  By the time an HCHII region has formed, the infalling region has grown. We therefore have a much higher infall detection rate because the infalling area is much better matched to our beam area.  For the UCHII regions, the infall detection rate drops to 50\%.  As shown above, the HCO$^+$ beam filling factor has drastically increased for these sources, and, the region undergoing infall should be easier to detect. That our detection rate is not as high may suggest that the outflows in the later stages have begun reversing the large scale infall, and some of these regions are reaching their final masses.

\subsection{Outflow}
\label{subsec:outflow}

SiO can be produced in either shocked gas from an outflow, or from a photodissociation region (PDR) surrounding, for instance, an HII region. For this survey we favor the outflow shock scenario  because of the high SiO detection rates in the HMPOs which have yet to form HII regions or create PDRs.  We note that the observations of \citet{Maxia01} suggest an outflow as the origin of the SiO enhancement in G31.41 and G29.96, which are also included in this study

As shown in Figure \ref{fig:frac_stage}, the fraction of sources in each stage with active outflow motions changes. Unpublished SiO (J=2-1) data from the IRAM 30m taken simultaneously with the CO data of \citet{Beuther02} of the HMPOs in this sample show detections in SiO (J=2-1) for 11/12 of the HMPOs in our study.  This is a 92\% detection rate (at a velocity resolution of $\sim$ 3.45 km s$^{-1}$, and an rms noise limit of approximately 0.03 K) whereas at the J=8-7 transition, our detection rate is slightly lower at 58\%.  \citet{Gusdorf08} calculate relative intensities of various rotational transitions of SiO, and find that the 2-1 transition is brighter than the 8-7 transition when the shock velocity is less than 30 km s$^{-1}$.  They also point out that for high density ($n\sim10^6$ cm$^{-1}$), shocks with velocities greater than $\sim$ 32 km s$^{-1}$ become dissociative J-shocks. As high mass star formation tends to happen in regions with densities this high, we suggest that shock velocities of $\sim$ 27 km s$^{-1}$ could account for our lowered SiO J=8-7 detection rate in our sources, regardless of evolutionary stage.

It is unclear why the detection rate of SiO drops below 50\% for the HCHII regions when, for both HMPOs and UCHII regions, the detection rates are much higher.  The integrated intensities of the SiO for HCHII regions with detections correlate very well with those found for the other stages. In fact, this is our tightest correlation in Figure \ref{fig:averages}.

The bottom left panel of Figure  \ref{fig:distance} shows the integrated intensity of SiO plotted against the full width at half maximum (FWHM) of the H$^{13}$CO$^+$ line. Along with this, we plot the logarithms of these values against the log of the distances to the sources.  We note that the intensity of the SiO line does not correlate with distance to source (correlation coefficient = 0.09), however the FWHM of the H$^{13}$CO$^+$ does (correlation coefficient = 0.55). Since the observing beam is the same for each H$^{13}$CO$^+$ observation, increased distance to source corresponds to a larger observed area. The correlation in the top right panel of Figure \ref{fig:distance} is approximately the Larson size-linewidth relation \citep{Larson81}. The bottom right panel of Figure \ref{fig:distance} shows a plot of SiO integrated intensity vs. the FWHM of H$^{13}$CO$^+$ corrected for the equation given in the top right panel (corrected for the size-linewidth relation).  The correlation between the integrated intensity of SiO and linewidth remains. Many of the line widths are still much greater than thermal as well suggesting that it is easier to detect SiO in more turbulent regions.  It is not clear whether this relationship is due to, or a product of  the shocks responsible for the SiO intensity; It is the integrated intensity of the SiO that scales with the FWHM, not just the detections of SiO.

\begin{figure}
\centering
\includegraphics[width=\columnwidth]{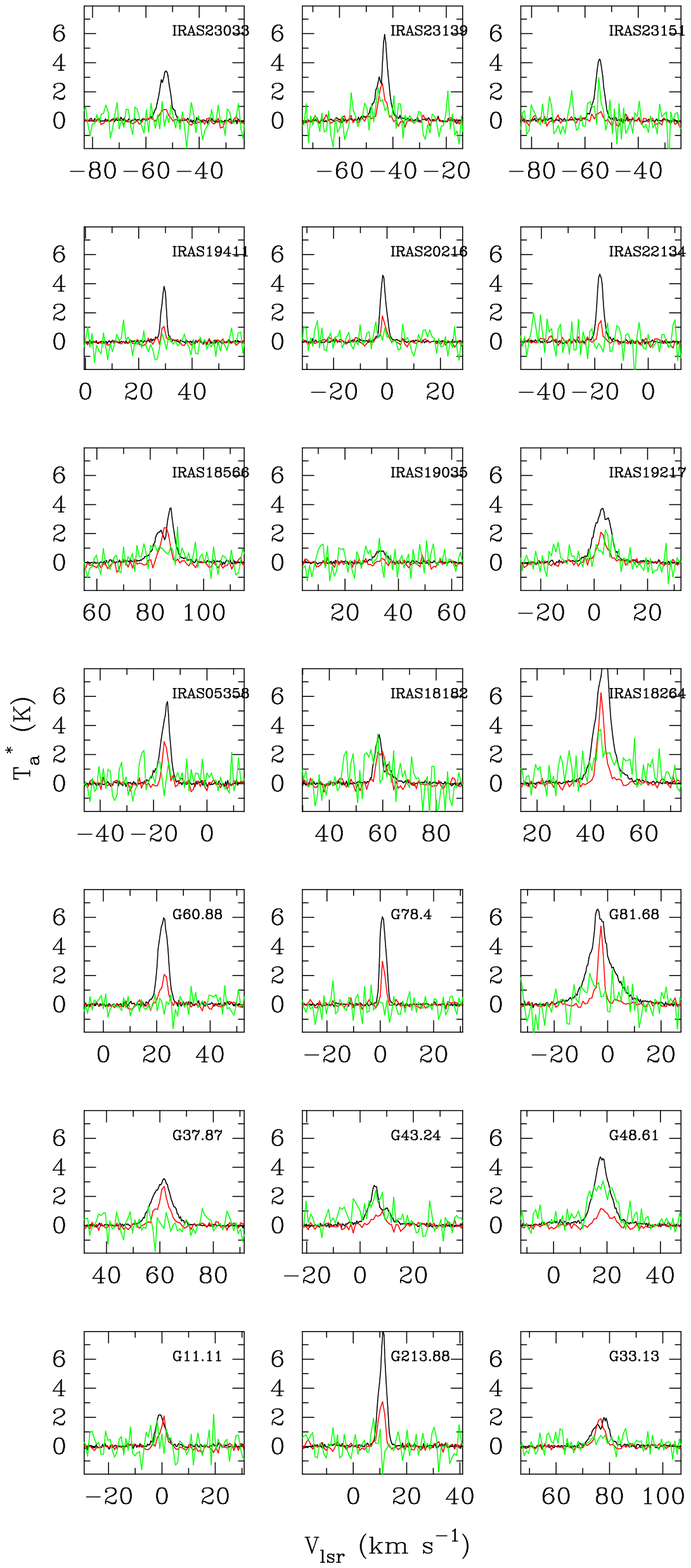}
\caption{Spectra from new sources. Spectra of previously published sources can be found in \citet{KW07} and \citet{KW08}. The scale on the Y axis reflects the intensity of the HCO$^+$ line(black). The H$^{13}$CO$^+$  (red) has been scaled up by a factor of four, and the SiO (green) has been scaled up by a factor of 16.}
\label{fig:spectra}
 \end{figure}

\begin{figure}
\centering
\includegraphics[angle=-90,width=\columnwidth]{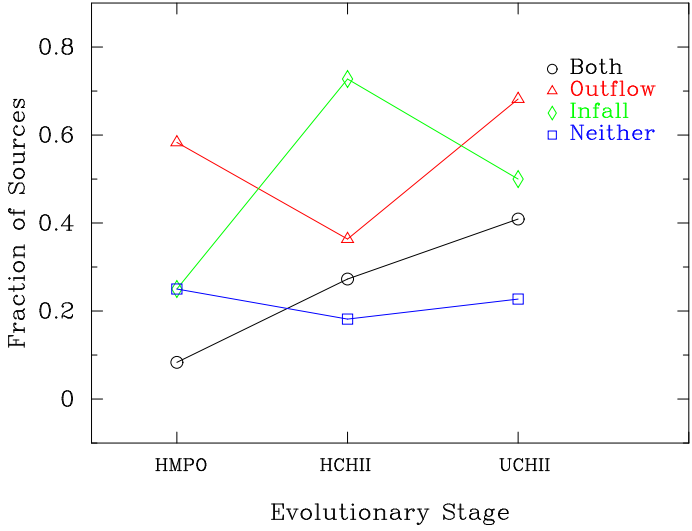}
\caption{Fraction of sources in each evolutionary bin with either infall ($\diamondsuit$), outflow ($\triangle$), both ($\circ$) or neither ($\Box$) signature detected. Here, any source with infall or outflow is plotted with their respective datapoint as well as with the `both' labels as appropriate.  Plotting this way shows that there is strong evidence for ongoing infall and outflow during the later stages of evolution.}
\label{fig:frac_stage}
 \end{figure}

\begin{figure}
\centering
\includegraphics[angle=-90,width=0.9\columnwidth]{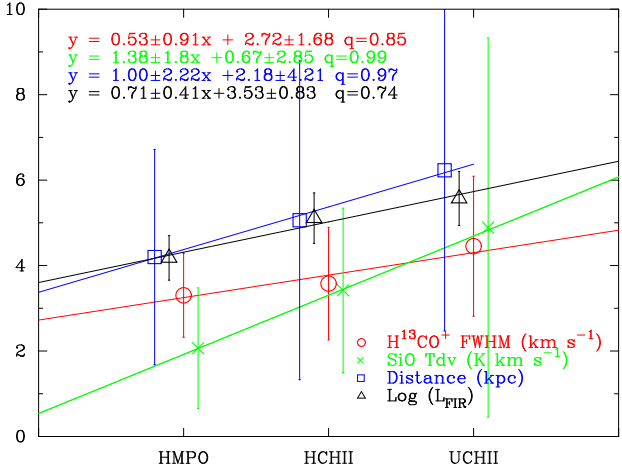}
\caption{Source properties averaged over evolutionary stage.  The correlation coefficients for the un-averaged properties are 0.025, 0.25, 0.71 and 0.35 for the H$^{13}$CO$^+$ FWHM, Distance, FIR Luminosity and SiO integrated intensity (respectively).}
\label{fig:averages}
\end{figure}

\begin{figure*}
\centering
\includegraphics[angle=-90,width=0.8\textwidth]{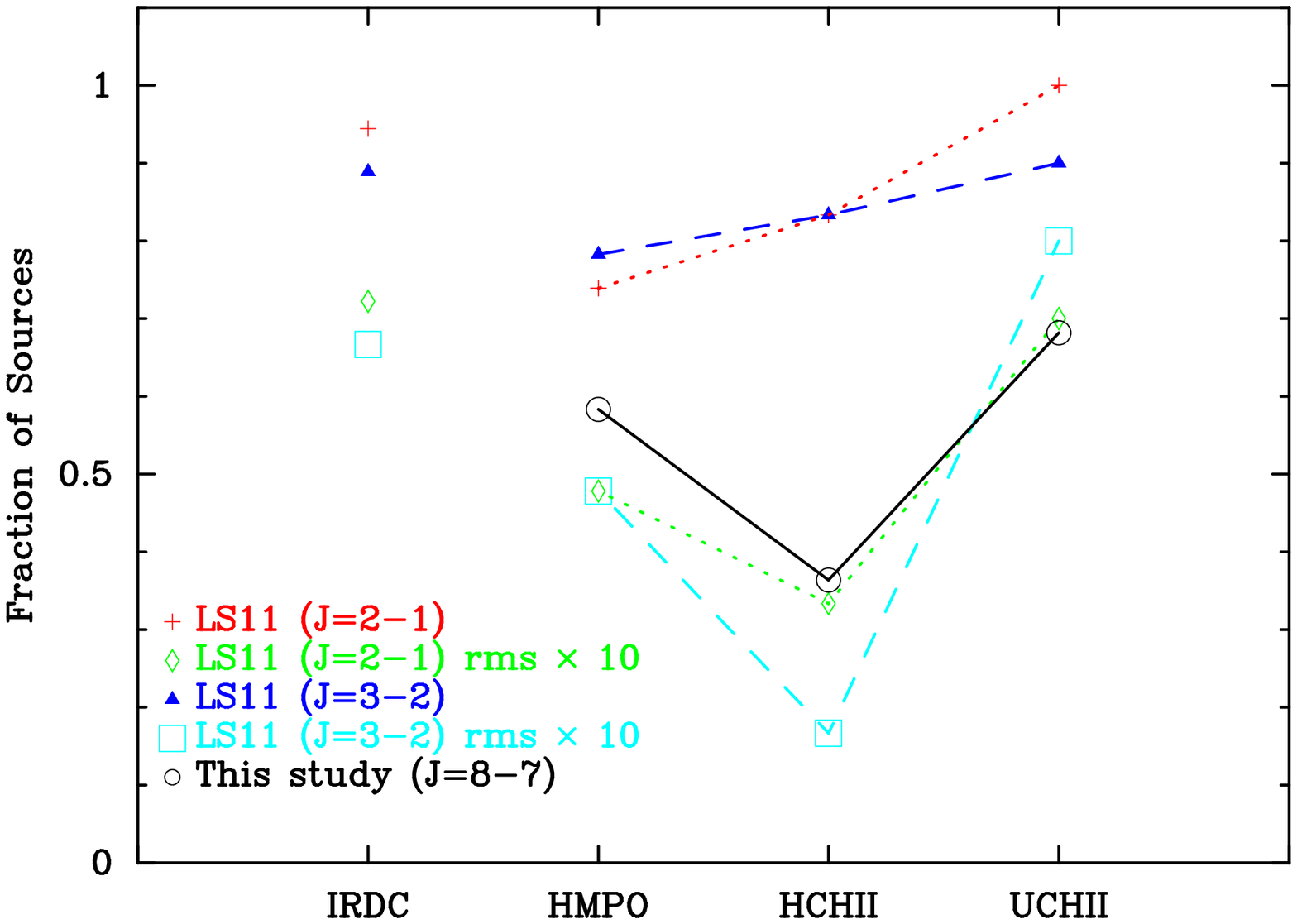}
\caption{Comparison between the detection rate of SiO presented here with that of \citet{LS11}.  The small dashed lines represent their J=2-1 detections, and the large dashed lines represent their J=3-2 detections.  The $\diamondsuit$ and $\Box$ symbols show the \citet{LS11} detections correspond to degrading their sensitivity by a factor of 10 to match our sensitivity.  Note the corresponding lack of HCHII region SiO detections in all three tracers. This plot does not take into account differences in beam sizes or excitation conditions required for the 3 different SiO transitions. \citet{LS11} contained sources at an earlier evolutionary stage than that probed here \citep[from the study of][]{Rathborne06} and those sources have been included here for completeness. }
\label{fig:comparison}
\end{figure*}

\begin{figure}
\centering
\includegraphics[angle=-90,width=0.9\columnwidth]{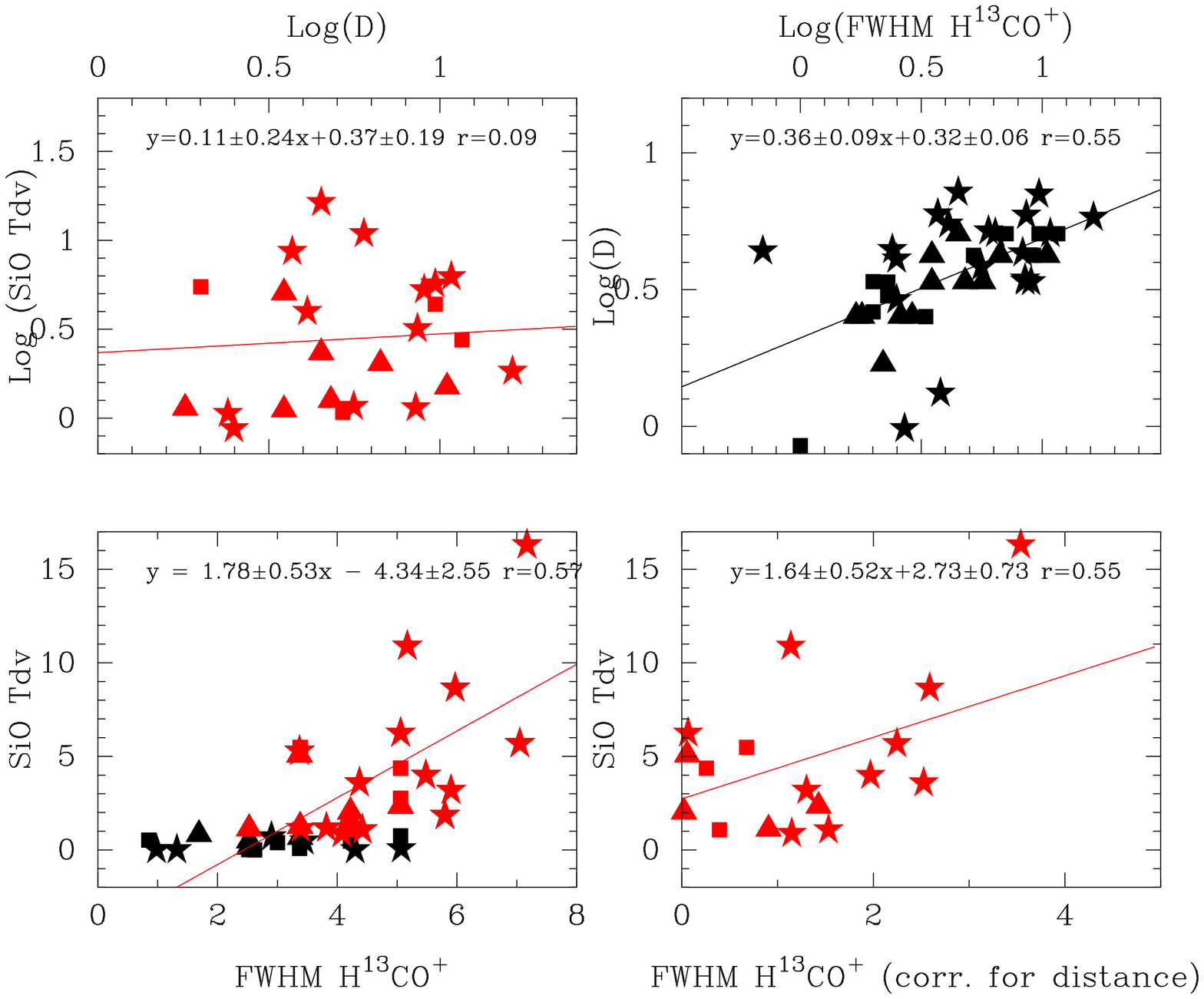}
\caption{(bottom left) SiO integrated intensity vs. H$^{13}$CO$^+$ FWHM and distance. Black symbols in for SiO plots  show non-detections, red points indicate detections. The triangles ($\blacktriangle$) indicate HMPOs, the filled squares ($\blacksquare$) indicate HCHII regions and the filled stars ($\bigstar$) indicate UCHII regions. The top panels show these two quantities as a function of source distance.  The bottom right panel shows the SiO integrated intensity vs. H$^{13}$CO$^+$ FWHM corrected for the size-linewidth relationship. }
\label{fig:distance}
\end{figure}


\section{Discussion \& Conclusions}
\label{sec:discussion}

Above we have noted the role beam dilution may have in our non-detections, however our results are, in general consistent with previous studies which have either looked at the infall or outflow signatures in massive star forming regions at different evolutionary stages. Our results are consistent with the region undergoing large scale infall growing with time.

We note the case of IRAS20126+4107. We do not detect SiO, yet there is a well known SiO jet coming from this source \citep{Cesaroni99}.  The intensity and size of this jet are small enough that we would expect it to be too beam diluted to be detected in our current survey. However, our infall detections are consistent with the the high resolution observations of \citet{Beltran11} for G10.6, G19.61 and G29.96. We, and they, detect infall signatures in the first two sources, but not in the third.  Their infall velocities, and thus mass infall rates are much higher than ours, which is due to different methods for determining the infall velocity (two layer infall modeling vs. peak velocity shifting).  \citet{Maxia01} suggest large scale infall is occurring in both G29.96 and G31.41.  The later is consistent with our results, the former is not.  

Our outflow detection rates appear to be at odds with the SiO results of \citet{LS11}, however this discrepancy appears to be due to our sensitivity limits.  In the UCHII region stage, there are no sources with infall signatures that do not show evidence for outflows. This suggests that if a source is still infalling at this late stage, it is likely that it is still powering a large outflow.  The older sources with SiO outflow signatures and no infall signatures may be showing the remnants of previous infall activity, as SiO only stays in the gas phase for of order  10$^4$ yr.  Our results are consistent with the region undergoing bulk infall growing with time.

 Although we see no clear infall vs. outflow ratio with age, we can say from our survey that as a high mass star forming region ages, the likelihood of observing both infall and outflow together increases (black line in Figure \ref{fig:frac_stage}), and that the intensity of the SiO that is detected also increases with the age of the outflow.

\begin{acknowledgements}
      The authors would like to thank Remo Tilanus for his help in reducing the data, and the anonymous referee and the editor for their insights which sharpened the arguments presented in this manuscript. LT acknowledges financial support from ASI under contract with INAF-Osservatorio Astrofisico di Arcetri.
\end{acknowledgements}

\bibliographystyle{aa}

\end{document}

%% file: tab1.tab
\begin{table*}
\caption{Observed Sample}             
\label{tab:observations}      
\centering          
\begin{tabular}{lllccccc}     
\hline\hline       
Name   &\multicolumn{2}{c}{Position (J2000)} & \multicolumn{2}{c}{RMS noise limit (K)}&V$_{\rm LSR}$&\multicolumn{2}{c}{Distance}\\
\cline{2-3} \cline{4-5} \cline{7-8} \\
&  RA &DEC&  347 GHz& 356 GHz	&(km s$^{-1}$)&(kpc) & ref\\
\hline\multicolumn{8}{c}{New Data}\\
\hline
G11.11-.4&18 11 32&-19 30 39& 0.049 & 0.097 &1&5.2&4\\
G213.88-11.84&06 10 51&-06 11 50& 0.049 & 0.101 &10&5.2&4\\
G033.13-00.09&18 52 08&+00 08 12& 0.041 & 0.073 &76.9&7.1&4\\
G037.87-00.40&19 01 54&+04 12 49& 0.040 & 0.074 &58.8&9.2&4\\
G043.24-00.05&19 10 34&+09 08 24& 0.040 & 0.076 &7.3&11.6&4\\
G048.61+00.02&19 20 31&+13 55 23& 0.041 & 0.090 &2.3&9.7&4\\
G060.88-00.13&19 46 20&+24 35 29& 0.042 & 0.102 &23.3&2.3&4\\
G78.438+2.659&20 19 39&+40 56 38& 0.048 & 0.096 &3.3&3.3&4\\
G081.68+00.54&20 39 01&+42 19 53& 0.045 & 0.102 &-8.2&2&4\\
IRAS05358+3543&05 39 15&+35 46 20& 0.047 & 0.115 &-16.5&1.8&3\\
IRAS18182-1433&18 21 10&-14 31 44& 0.060 & 0.091 &59&4.5&3\\
IRAS18264-1152&18 29 15&-11 50 18& 0.045 & 0.088 &44.5&3.5&3\\
IRAS18566+0408&18 59 10&+04 12 14& 0.060 & 0.074 &86.5&6.7&3\\
IRAS19035+0641&19 06 02&+06 46 39& 0.039 & 0.077 &32.5&2.2&3\\
IRAS19217+1651&19 23 59&+16 57 50& 0.041 & 0.089 &3&10.5&3\\
IRAS19411+2306&19 43 18&+23 13 59& 0.042 & 0.090 &30&2.9&3\\
IRAS20126+4107&20 23 24&+41 17 36& 0.050 & 0.068 &-1.5&1.7&3\\
IRAS22134+5834&22 15 09&+58 49 09& 0.050 & 0.072 &18.5&2.6&3\\
IRAS23033+5951&23 05 25&+60 08 08& 0.052 & 0.076 &-53.5&3.5&3\\
IRAS23139+5939&23 16 11&+59 55 33& 0.048 & 0.080 &-43&4.8&3\\
IRAS23151+5912&23 17 21&+59 28 49& 0.061 & 0.071 &-56&5.7&3\\
\hline
\multicolumn{8}{c}{\citet{KW08} - maps}\\
\hline
G5.89-0.39	&18 00 30.5      &-24 03 58.00&0.238&0.070&9&2&2\\
G10.47+0.03	&18 08 38.4      &-19 51 52.00&0.082&0.074&67&10.8&2\\
G10.6-0.4		&18 10 28.0      &-19 56 04.00&0.102&0.079&-3&6&2\\
G19.61-0.23	&18 27 38.3      &-11 56 40.00&0.090&0.098&43&4.5&2\\
G20.08-0.14	&18 28 10.7      &-11 28 48.00&0.084&0.071&42&4.1&2\\
G29.96-0.02	&18 46 04.1      &-02 39 22.00&0.126&0.144&98&9&2\\
K3-50A		&20 01 45.8 	&+33 32 42.00&0.089&0.066&0.24&8.6&2\\
\hline
\multicolumn{8}{c}{\citet{KW07} - single pointing}\\
\hline
G5.97&18 03 40.4&-24 22 44&0.044&0.069&10&2.7&1\\
G8.67&18 06 19.0&-21 37 32&0.068&0.074&36&8.5&1\\
G12.21&18 12 39.7&-18 24 21&0.042&0.076&24&16.3&1\\
M17S&18 20 24.8&-16 11 35&0.044&0.077&20&2.3&1\\
G31.41&18 47 33.0&-01 12 36&0.044&0.061&97&8.5&1\\
G34.26&18 53 18.5&01 14 58&0.047&0.131&58&3.7&1\\
G45.07-0.13&19 13 22.1&10 50 53&0.044&0.073&59&9.7&1\\
G45.47&19 14 25.6&11 09 26&0.037&0.063&58&8.3&1\\
G61.48&19 46 49.2&25 12 48 &0.037&0.048&12&2&1\\
G75.78&20 21 44.1&37 26 40&0.038&0.068&0&5.6&1\\
Cep A&22 56 17.9&62 01 49 &0.031&0.076&-10&0.7&1\\
W3(OH)&02 27 03.8&61 52 25 &0.026&0.079&-48&2.4&1\\
G138.3&03 01 29.2&60 29 12&0.066&0.073&-38&3.8&1\\
G139.9&03 07 23.9&58 30 53&0.058&0.071&-39&4.2&1\\
G192.58&06 12 53.6&17 59 27 &0.037&0.053&9&2.5&1\\
G192.6&06 12 53.6&18 00 26&0.074&0.098&9&2.5&1\\
G240.3&07 44 51.9&-24 07 40&0.035&0.048&68&6.4&1\\
\hline                  
\end{tabular}
\tablefoot{Distances are given in: (1) \citet{KW07}, (2)\citet{KW08}, (3) \citet{Sridharan02}, and (4) \citet{Kurtz94}}
\end{table*}

%% file: tab2.tab
\begin{table}
\caption{Infall and Outflow Detections}             
\label{tab:detections}      
\centering          
\begin{tabular}{lccccc}     
\hline\hline     
Source  & Type  & Infall & Outflow & Offset &Notes\\
\hline
IRAS05358+3543		&		HMPO	&	N	&	CM	&	both&2-1\\
IRAS18182-1433	&	HMPO	&	CM	&	CM	&	both&2-1\\
IRAS18264-1152	&	HMPO	&	N	&	CM	&	&2-1\\
IRAS18566+0408	&	HMPO	&	N	&	CM	&	&2-1\\
IRAS19035+0641	&	HMPO	&	N	&	N	&	both&2-1\\
IRAS19217+1651	&	HMPO	&	N	&	CM	&	&2-1\\
IRAS19411+2306	&	HMPO	&	N	&	N	&	&2-1\\
IRAS20126+4107	&	HMPO	&	N	&	N	&	356&2-1\\
IRAS22134+5834	&	HMPO	&	CM	&	N	&	both&\\
IRAS23033+5951	&	HMPO	&	N	&	CM	&	both&2-1\\
IRAS23139+5939	&	HMPO	&	N	&	M	&	&2-1\\
IRAS23151+5912	&	HMPO	&	CM	&	N	&	347&2-1\\
\hline
G11.11-0.4	&	HCHII	&	CM	&	N	&	\\
G213.88-11.84	&	HCHII	&	M	&	M	&	both\\
G033.13-00.09	&	HCHII	&	N	&	N	&	\\
G037.87-00.40	&	HCHII	&	C	&	N	&	\\
G043.24-00.05	&	HCHII	&	CM	&	CM	&	\\
G048.61+00.02	&	HCHII	&	N	&	CM	&	\\
G060.88-00.13	&	HCHII	&	CM	&	N	&	both\\
G78.438+2.659	&	HCHII	&	M	&	N	&	347\\
G081.68+00.54	&	HCHII	&	CM	&	M	&	both\\
M17S	&	HCHII	&	Y	&	N	&	&1p\\
G61.48	&	HCHII	&	N	&	N	&	&1p\\
\hline
G5.89	&	UCHII	&	N	&	CM	&	\\
G10.47	&	UCHII	&	CM	&	CM	&	both\\
G10.6	&	UCHII	&	CM	&	CM	&	both\\
G19.61	&	UCHII	&	CM	&	CM	&	both\\
G20.08	&	UCHII	&	CM	&	M	&	both\\
G29.962	&	UCHII	&	N	&	CM	&	both\\
K3-50A	&	UCHII	&	N	&	CM	&	both\\
G5.97	&	UCHII	&	N	&	N	&	&1p\\
G8.67	&	UCHII	&	Y	&	Y	&	&1p\\
G12.21	&	UCHII	&	N	&	Y	&	&1p\\
G31.41	&	UCHII	&	Y	&	N	&	&1p\\
G34.26	&	UCHII	&	Y	&	Y	&	&1p\\
G45.073	&	UCHII	&	N	&	Y	&	&1p\\
G45.47	&	UCHII	&	N	&	Y	&	&1p\\
G75.78	&	UCHII	&	N	&	Y	&	&1p\\
Cep A	&	UCHII	&	Y	&	Y	&	&1p\\
W3(OH)	&	UCHII	&	Y	&	Y	&	&1p\\
G138.3	&	UCHII	&	N	&	N	&	&1p\\
G139.9	&	UCHII	&	N	&	N	&	&1p\\
G192.58	&	UCHII	&	Y	&	Y	&	&1p\\
G192.6	&	UCHII	&	Y	&	N	&	&1p\\
G240.3	&	UCHII	&	Y	&	N	&	&1p\\
\hline
\end{tabular}
\tablefoot{For those sources with offsets between the continuum and molecular peaks, C indicates a detection at the continuum peak, and M indicates a detection at the molecular peak. Column 5 indicates whether there was a shift between the continuum peak and the molecular peak.  (356) indicates an offset in the 356 GHz observations, (347) indicates an offset in the 347 GHz observations, and (both) indicates an offset in both tunings. Column 6 notes:  (1p) - Single pointing observations. (2-1) - SiO J=2-1 detection for the HMPOs}
\end{table}

%% file: tab3.tab
\begin{table}
\caption{Infall Rates}             
\label{tab:infall_vel}      
\centering          
\begin{tabular}{ll@{$\pm$}rl@{$\pm$}rl@{$\pm$}rl@{$\pm$}r}     
\hline\hline     
Source & \multicolumn{4}{c}{Continuum Peak} & \multicolumn{4}{c}{Molecular Peak}\\
& \multicolumn{2}{c}{V$_{\rm in}$} & \multicolumn{2}{c}{$\dot{\rm M}_{\rm in}$ }
& \multicolumn{2}{c}{V$_{\rm in}$} & \multicolumn{2}{c}{$\dot{\rm M}_{\rm in}$ }\\
& \multicolumn{2}{c}{{\bf km s$^{-1}$}} & \multicolumn{2}{c}{{\bf $10^{-2}\dot{\rm M}$yr$^{-1}$ }}
& \multicolumn{2}{c}{{\bf km s$^{-1}$}} & \multicolumn{2}{c}{{\bf $10^{-2}\dot{\rm M}$yr$^{-1}$}}\\
\hline
IRAS18182 	& 1.74 & 0.58 & 1.62 & 0.54 & 1.57 & 0.56 & 1.46 & 0.52\\
\hline
G43.24		& 1.27 & 0.33 & 5.62 & 1.44 & 1.61 & 0.45 & 7.12 & 2.01\\
G81.68		&\multicolumn{4}{c}{Shoulder} & 0.26 &0.07 &0.036 &0.01\\
\hline
G10.47      &\multicolumn{4}{c}{same as mol. peak}&  0.42 & 0.11 &  1.6 & 0.43\\
G10.6        & \multicolumn{4}{c}{same as mol. peak}&0.05 & 0.03  & 0.06 & 0.03\\
G19.61      & \multicolumn{4}{c}{same as mol. peak}& 0.21 & 0.08 &  0.14 & 0.05\\
G20.08      & \multicolumn{4}{c}{same as mol. peak}& 0.05 & 0.03 &  0.03& 0.02\\
G8.67	&0.4	&0.1		&0.04&0.02&\multicolumn{4}{c}{Single Pointing}\\
M17S	&1.4	&0.5		&0.04 & 0.02&\multicolumn{4}{c}{Single Pointing}\\
Cep A	&0.23&0.07	&0.002&0.001&\multicolumn{4}{c}{Single Pointing}\\
W3(OH)	&0.06	&0.02		&0.03&0.01&\multicolumn{4}{c}{Single Pointing}\\
G192.58	&0.8	&0.3	&\multicolumn{2}{c}{no data}&\multicolumn{4}{c}{Single Pointing}\\
G192.6	&0.9	&0.4		&0.02&0.01&\multicolumn{4}{c}{Single Pointing}\\
G34.26	&1.5	&0.3		&0.14&0.04&\multicolumn{4}{c}{Single Pointing}\\

\end{tabular}
\tablefoot{Single lines separate the three stages (HMPO, HCHII and UCHII regions from top to bottom). Velocities are in units of km s$^{-1}$ and mass infall rates are in units of $10^{-2}$ M$_\odot$ yr$^{-1}$.  Infall velocities derived for UCHII regions are given in \citet{KW07,KW08}.}
\end{table}

%% file: tab5.tab
\begin{table}
\caption{Infall and Outflow Statistics}
\label{tab:stats}
\centering
\begin{tabular}{lccccc}
\hline \hline

Stage & \# of sources & Infall & Outflow & Both & Neither\\
\hline
HMPO 	& 12 &3	&7 	&1 	& 4\\
HCHII 	& 11 &8	&4 	& 3	& 2\\
UCHII 	& 22 &12	&16 & 9	& 5\\
\hline
\hline
\end{tabular}
\tablefoot{Fraction of infall and outflow detections as a function of evolutionary stage.}
\end{table}

%% file: tab4.tab
\begin{table}
\caption{Source Property Averages by Evolutionary Stage}
\label{tab:averages}
\centering
\begin{tabular}{rl@{$\pm$}ll@{$\pm$}rl@{$\pm$}rl@{$\pm$}r}
\hline \hline
Stage &\multicolumn{2}{c}{H$^{13}$CO$^+$ FWHM} & \multicolumn{2}{c}{SiO $\int Tdv$} & \multicolumn{2}{c}{Distance} & \multicolumn{2}{c}{Log(L$_{\rm FIR}$)}\\
& \multicolumn{2}{c}{(km s$^{-1}$)}&\multicolumn{2}{c}{(K km s$^{-1}$)}&\multicolumn{2}{c}{(kpc)}\\
\hline
HMPO      & 3.305 &  0.982&   2.067 &  1.410&   4.20 &   2.52&    4.18    &0.52\\
HCHII       &3.578  & 1.316  & 3.422   &1.922  & 5.06&    3.73  &  5.11   & 0.59\\
UCHII      &4.454   &1.636   &4.890   &4.436   &6.23&    3.76    &5.57  &  0.62\\
\hline
slope &0.5&0.9 &1.4&1.8 &1&2 &0.7&0.4\\
q\tablefootmark{a} & \multicolumn{2}{c}{0.85}& \multicolumn{2}{c}{0.99}& \multicolumn{2}{c}{0.97}& \multicolumn{2}{c}{0.74} \\ 
\hline
\hline
\end{tabular}
\tablefoot{Source properties averaged over evolutionary stage. The unceratinties reflect the standard deviation in the averaged quantities.}
\tablefoottext{a}{Correlation coefficient}
\label{tab:averages}
\end{table}